 \documentstyle[epsf,twocolumn,aps,graphics,floats]{revtex}
\begin{document}
\draft

\flushbottom 
\twocolumn[\hsize\textwidth\columnwidth\hsize\csname@twocolumnfalse\endcsname

\title{\bf First principles study of the adsorption of
C$_{60}$ on Si(111)}

\author{
Daniel~S\'anchez-Portal$^{1}$,
Emilio~Artacho$^2$,
J.~I.~Pascual$^{2,3}$,
J.~G\'omez-Herrero$^2$, Richard M. Martin$^1$,
and Jos\'e~M.~Soler$^{2,4}$
}

\address{
$^1$Department of Physics and Material Research Laboratory, \\
University of Illinois, Urbana, IL-61801,USA \\
$^2$Departamento de F\'{\i}sica de la Materia Condensada, C-III,
and Instituto Nicol\'as Cabrera, \\
Universidad Aut\'onoma de Madrid, 28049 Madrid, Spain\\
$^3$ Fritz-Haber-Institut der Max-Planck Gesellschaft,
Faradayweg 4-6, D-14195 Berlin, Germany \\
$^4$ Department of Physics, Lyman Laboratory, Harvard University,
Cambridge, MA 02138, USA
}

\date{ECOSS-19 abstract 00673}

\maketitle

\begin{abstract}
   The adsorption of C$_{60}$ on Si(111)
has been studied by means of
first-principles density functional calculations.
   A 2x2 adatom surface reconstruction was used to simulate
the terraces of the 7x7 reconstruction.
   The structure of several possible adsorption configurations
was optimized using the {\it ab initio} atomic forces,
finding good candidates for two different adsorption states
observed experimentally.
   While the C$_{60}$ molecule remains closely spherical,
the silicon substrate appears quite soft, especially the
adatoms, which move substantially to form extra C-Si bonds,
at the expense of breaking Si-Si bonds.
   The structural relaxation has a much larger effect on the
adsorption energies, which strongly depend on the
adsorption configuration, than on the charge transfer.
\end{abstract}

\pacs{Keywords: Density functional calculations, Fullerenes,
Chemisorption}

]

Recent studies using scanning tunneling microscopy (STM) have reported
images that reveal intra-molecular features of C$_{60}$ molecules
deposited, 
at very low coverages, on
Si(100)-(2$\times$1)~\cite{Yao-96}, and
Si(111)-(7$\times$7)~\cite{Hou,nacho,nacho-prl}. 
These observations 
are very interesting as they might
be used to infer the orientation of the molecules~\cite{Hou}, 
and other 
details of the bonding configuration. They also confirm 
the strong interaction of the fullerenes with these surfaces.
Furthermore, in the images reported by Pascual {\em et al.}~\cite{nacho} 
it is 
clear that there exist two types of molecules: ``large" molecules with
an apparent height of $\sim$0.6~nm and width of $\sim$2.0~nm, which
appear more
round and fuzzy in the constant current STM images, 
and ``small" molecules with 
a height of $\sim$0.5~nm and width of $\sim$1.5~nm, 
which present a more clearly discernible
internal structure and a larger variety of shapes. 
A similar observation of two different adsorption states has also
been done by 
Yao {\em et al}~\cite{Yao-96} on the Si(100)-(2$\times$1) substrate.
Interestingly enough, 
after annealing at 870~K the ``large" molecules
evolve to ``small" ones, indicating
that they were probably in a weaker adsorption state. 

In this paper we apply first-principles
electronic structure methods to study the adsorption 
of fullerenes on the terraces of the Si(111)-(7$\times$7)
reconstruction, identifying
candidates for the two adsorption states observed. There are some
previous electronic structure calculations of fullerenes over silicon
substrates~\cite{images-1,Yamaguchi,Hou}.
However, 
in the sole case where a structural optimization
was performed~\cite{Yamaguchi}, a semi-empirical force field
was used. 
Therefore, to the best 
of our knowledge this is the first systematic study of 
the adsorption geometry of C$_{60}$ on a silicon surface
using {\it ab initio} atomic forces.

The calculations have been performed with the SIESTA 
program~\cite{SIESTA,Emilio}, which allows standard calculations 
within density functional theory~\cite{Kohn-Sham} (DFT)
for systems with hundreds of atoms.
It uses norm-conserving pseudopotentials~\cite{Troullier-Martins},
and a basis set of numerical atomic orbitals, obtained from the 
solution of the atomic pseudopotentials at a slightly excited 
energy~\cite{Emilio,Sankey-Niklewski}.\footnote{ We have used here an 
`energy shift' of
0.5~eV. The corresponding radii are 4.65  and 5.73~a.u. for the $s$ and
$p$ orbitals of Si, 3.79 and 4.41 in the case of C, and 4.26 for the $s$
states of H. }
In this work we have used the local 
density approximation to DFT~\cite{LDA} and a minimal $sp^3$ 
basis set for both C and Si.
With this basis,
the Si-Si and C-C bond lengths differ only 1\% and 2\% from experiment,
respectively. Our method also uses 
a real-space grid to compute the Hartree and the exchange correlation 
contributions to the self-consistent potential and the
Hamiltonian matrix.
Real and reciprocal-space integrations were performed with a
70~Ry-cutoff grid and with 2 inequivalent k-points.
For a given basis set,
these meshes guarantee a convergence better than $\sim$ 1~meV/atom
for the total energies and of $\sim$0.1~eV 
for the C$_{60}$ binding energies~\footnote{
The binding energies have been corrected for the
basis set superposition error as described, for example,
in Ref.~\cite{BSSE}}

The terraces of the Si(111)-(7$\times$7) surface
have been  modeled by slabs of two double layers, 
covered with a 2$\times$2 adatom reconstruction on one face, 
and saturated with hydrogen on the other face.
The C$_{60}$ molecules are arranged in a $2\sqrt{3}\times 2\sqrt{3}$
periodic supercell, big enough to avoid strong interactions between
them.
First, we relax the structure of the clean surface 
with the lattice constant parallel to the surface fixed at the 
calculated bulk value, to avoid artificial stresses.
The relaxed structure is 
similar to that of other authors~\cite{Vanderbilt}.
The distance between the pedestal atoms (restatoms bound to the adatoms) 
is reduced, 
and a strong downward relaxation
for the atoms beneath the adatom is observed. The rest atom lies
$\sim$0.4~\AA\ higher than the ideal bulk height. The main discrepancy with
the results of Ref.~\cite{Vanderbilt} is that the position of the
adatoms
is approximately 9\% higher in our case. This difference may be due to
the use of a minimal basis. However, this is not crucial, since we are
interested in the changes induced by the adsorption of the C$_{60}$,
rather than in the surface itself.
Our results also reproduce
the main features of the electronic structure of the surface.

\begin{figure}
\epsfxsize=8.0cm\centerline{\epsffile{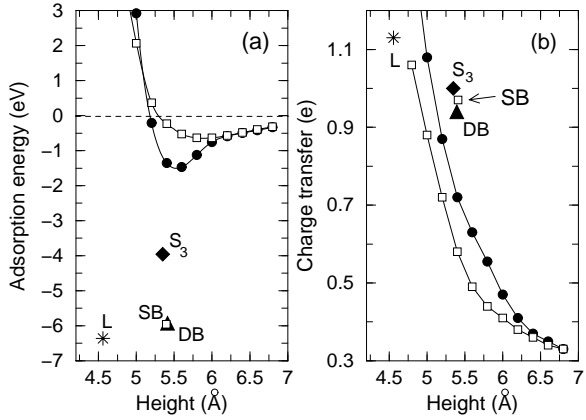}}
\caption[]{
a) Binding energies of the C$_{60}$ molecule on Si(111)-(2$\times$2),
for different orientations, as a function of the height of the
molecular center of mass over the initial position of the rest atom.
b) Like (a), but for
the charge transfer.
Open squares stand for the threefold-oriented unrelaxed structure,
and filled circles for the bond-oriented unrelaxed configuration.
The relaxed structures are labelled following the text.}
\label{energies}
\end{figure}

   First we place the C$_{60}$ molecule on top of one of the rest atoms, 
with different orientations, suggested by the STM images of 
Pascual {\em et al.}~\cite{nacho}.
   In one of them (labelled S$_3$) the molecule is oriented with an hexagon 
on top, to preserve the threefold symmetry of the substrate. 
In the other orientation, a C$_{60}$ double-bond (DB)
is placed directly over the rest atom.
   Fig~\ref{energies}(a), shows the binding energy as a function of the 
height of the center of mass of the fullerene relative to the initial
position of the rest atom.
   If no internal relaxation of the molecule or the surface are allowed,
the binding energies are 0.64~eV and 1.47~eV for the S$_3$ and DB
configurations, respectively.
   The DB molecule is more bound because it is lower and 
closer to the rest atom. 

\begin{figure*}
\epsfxsize=13.0cm\centerline{\epsffile{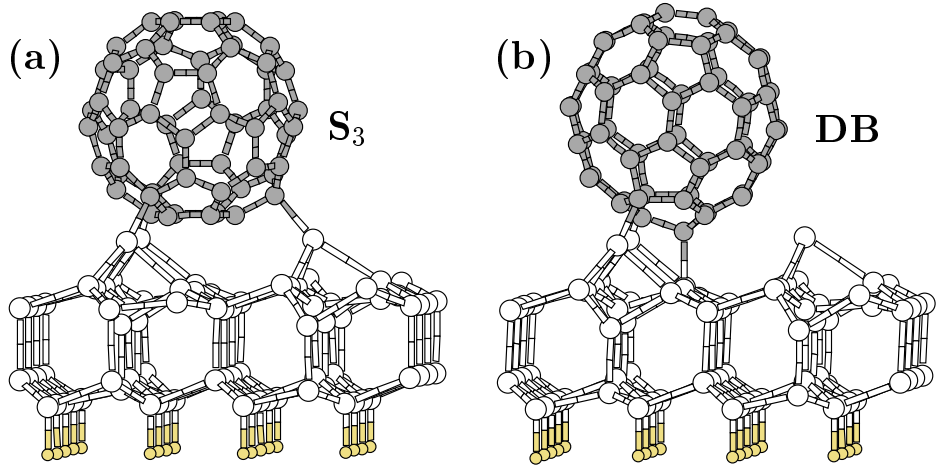}}
\caption[]{
Two different configurations of the C$_{60}$ molecule adsorbed on
the Si(111)-(2$\times$2) surface. a) Structure labelled as S$_3$
in the text,
the molecule preserves the three fold symmetry of the substrate,
presenting an hexagon on the top. b) DB structure
which, after the relaxation, has formed
a bond with the rest atom.
}
\label{structures1}
\end{figure*}

   We then investigate the structural changes, shown in
Fig~\ref{structures1},
in both the surface and the molecule after the adsorption, relaxing the 
atomic positions according to the {\it ab initio} forces. 
The carbon atoms move very little from their ideal positions, the
molecule 
remains almost spherical, 
and its height over the substrate decreases
from $\sim$5.9 to 5.35~\AA. 
   In contrast, the three nearest adatoms to the S$_3$ molecule move 
substantially, approaching the closest carbon atoms from 
$\sim$3.0~\AA\ to 2.02~\AA. 
   The rest atom below the molecule relaxes downward 0.61~\AA, while the 
pedestal atoms closer to the molecule relax upwards 0.08~\AA, and their 
bond distance to the adatom decreases from 2.57~\AA\ to 2.51~\AA. 
   The pedestal atoms more distant to the molecule relax upwards
0.24~\AA, 
and their bond length to the adatoms becomes 2.75~\AA. 
   The relaxation energy is large, increasing the binding energy to
3.96~eV.

   The case of the DB molecule in rather different. 
   It rotates during the relaxation, binding to the rest atom 
and to two adatoms.
   The bond length with the adatoms is 2.01~\AA, while the bond with the
rest atom is a little shorter (1.98~\AA) and stronger (larger bond
charges).
   The relaxation of these atoms, and those bound to them, is 
similar to that for the S$_3$ molecule, but the displacements are
somewhat smaller. 
   The deformation of the molecule is again very small.
   The third adatom is largely displaced towards the C$_{60}$, but the 
large distance to the molecule (2.63~\AA), and the very low bond charge 
do not indicate the formation of covalent bonding. 
   Its displacement is probably driven by the movement
of the pedestal atoms, and the ionic interaction with the
negatively charged
molecule.
   The equilibrium height is reduced from $\sim$5.5 to 5.39~\AA.
   The DB molecule has a binding energy of 5.96~eV, quite larger than 
the S$_3$ molecule. 
   This difference is due to the stronger bonds formed
and to the somewhat less strained surface.

   Fig.~\ref{energies} shows data for two other relaxed structures. 
   The structure labelled as SB (not shown) has initially a single-bond 
pointing towards the rest atom. 
   Like the DB structure, it forms, after the relaxation, 
a strong convalent bond with the rest atom.
   However, in this case the adatom more distant to the molecule
suffers a larger displacement, breaking its bond 
with one of the pedestal atoms, and forming another covalent bond 
with the fullerene. 
   This fourth bond is longer (2.31~\AA) and weaker than the other
three. 
   In fact, in spite of the formation of this extra-bond, the binding 
energy (5.93~eV) and its height (5.41~\AA) are almost identical to 
those of the DB molecule. 
   This indicates that the adatom most loosely bound to the molecule 
can probably oscillate between two positions with the same energy: 
one where it is covalently bound to the fullerene,
and another where the interaction with the molecule is mainly ionic.

\begin{figure}
\epsfxsize=8.0cm\centerline{\epsffile{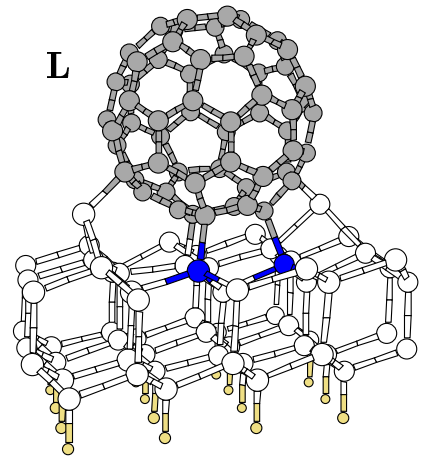}}
\caption[]{
Structure of the most stable adsorption configuration found
in our calculations (labelled L).
Two of the adatoms have broken their bonds with the correspoding
pedestal atom (highlighted in the figure),
allowing the formation of two extra bonds with the
C$_{60}$ molecule. }
\label{structures2}
\end{figure}

   The S$_3$ configuration might be a good candidate for the ``large" 
(less stable) molecules observed by STM~\cite{nacho}, whose resolved 
internal structure present a triangular shape~\cite{nacho}, 
consistent with the threefold symmetry.
   The DB and SB configurations are more bound to the substrate, 
but their height is almost identical to that of S$_3$.
   Although STM does not directly measure the atomic positions,
our estimations, using the Tersoff-Hamann~\cite{Tersoff-Hamann} theory, 
also indicate a similar ``electronic height" for the 
DB and S$_3$ configurations ($\sim$6.8\AA).
   A reasonable candidate for the ``small" molecules should have a 
higher binding energy and a lower heigth over the substrate than the 
structures studied to this point. 
   The relaxed structure labelled L, which fulfills these conditions,
is shown in Fig.~\ref{structures2}. This is just an example of the
several configurations, with different molecular orientations,
that can be obtained by allowing a larger
rearragement of the substrate than the structures previously studied.
These new structures are characterized by a lower height, and the 
formation of more bonds
between the molecule and the surface atoms.
 The L configuration
 has a binding energy of 6.36~eV (0.4~eV larger than the DB), its 
height is 4.56~\AA\  ($\sim$0.8~\AA\ lower than the other
configurations).
   The molecule presents an hexagon on the top and occupies almost a 
``bridge'' position between two adatoms.
   Each of these adatoms has a broken bond with the substrate, 
allowing for the formation of an extra-bond with the molecule 
(the new bonds have been highlighted in Fig.~\ref{structures2}). 
   The molecule has five
bonds with the substrate, two with the adatoms (2.13~\AA),
two with the freed pedestal atoms (1.98 and 2.01~\AA), 
and one with the rest atom (1.96~\AA). 

Our findings indicate that adsorption states like S$_3$, DB, or SB,
can be accessed whitout overcoming any energetic barrier, 
in other
words we have not found any physisorbed precursor for the adsorption of
C$_{60}$ over Si(111). However, 
it must exist an energetic barrier between the ``large" and 
``small" molecules 
observed experimentally, 
so both can be simultaneously visualized
at room temperature. The nature of the barrier has been 
clarified by our calculations: some of the bonds of the adatoms
with the substrate have to be broken to allow the formation
of extra-bonds with the molecule. Only after annealing 
the sample the majority of the molecules are
found in the lower configuration.
   It is also worth noting here that some
recent evidence supports a strong rearrangement of the Si(111) 
surface upon the adsorption of C$_{60}$~\cite{Sakamoto-99}. 
   In these experiments, after annealing one monolayer of C$_{60}$ at
670~K,
the (7$\times$7) diffraction pattern is lost and a (1$\times$1) pattern
appears in its place. 
   This transition is accompanied by a change 
to a more strongly bound adsorption state of the molecules.

The values of the charge transfer, from the Si(111)-(7$\times$7)
substrate to the C$_{60}$,
deduced from different experimental techniques, oscillate between
3$\pm$1 electrons~\cite{Suto-97}, and $\sim$0.21
electrons~\cite{Sakamoto-99}.
The theoretical
calculations might be important to understand this discrepancy.
The charge transfers
reported here are obtained by Mulliken analysis~\cite{Mulliken}.
Fig.~\ref{energies}(b) shows the evolution of the charge transferred
to the molecule  versus the distance to the rest atom. The charging of
the
molecule increases rapidly as the molecule approaches the surface.
For the relaxed structures the charge transfers are: 1.00 electrons for
the 
S$_3$, 0.94 and 0.97 for the DB and SB molecules, and 1.13 for the L
configuration.
Most of the transferred charge is in the atoms directly bound to the
surface:
0.16-0.17 electrons in the
carbon atoms bound to the rest atom, 0.12-0.14 electrons in those bound
to 
the adatoms, and 0.16 in those bound to the pedestal atoms in the L
molecule.
The rest of the charge, $\sim$0.5 electrons in all the cases, is mainly 
in the atoms closer to the surface, rather than uniformly distributed. 
This is an indication of the nature of the charge 
transfer. The charging of the molecule does not come through the
occupation
of the initially unoccupied states of the 
fullerene, but through the hybridization with the surface states. This
result
agrees with recent experimental 
reports of the transport through 
C$_{60}$ on Si(111).~\cite{nacho}.

   A previous calculation with the DV-X$_\alpha$-LCAO 
method~\cite{Yamaguchi}, has reported a considerably larger charge
transfer 
of 3.35 electrons.
   We believe that this discrepancy stems from the inherent
arbitraryness
of the Mulliken analysis to split the charge to covalently bound atoms, 
which makes the results strongly dependent on the basis set used. 
   However, we stress that both calculations agree on the general
pattern of the charge transfer.

   In conclusion, 
the adsorption of C$_{60}$ on Si(111)-(7$\times$7) has been studied
by 
{\it ab initio} density functional calculations. 
   Different adsorption configurations
have been explored, finding good candidates for the two adsorption 
states experimentally observed. 
   The adsorption energies range between
4 and 6.5~eV, while the charge transfer is always very close to one 
electron and mainly localized on the carbon atoms bound to the
substrate. 
Work is in progress to simulate the STM images of the different
adsorption structure described in this paper. We hope this will
allow a detailed comparison with the experiments. Molecular
dynamics simulations are also in progress to explore the mechanism,
and energetic barriers for the transition between different
adsorption configurations.

This work was supported by 
Grants No. DOE~8371494,
and No. DEFG~02/96/ER~45439.

\end{document}